\documentclass[twocolumn,epjc3]{svjour3}  

\smartqed  % flush right qed marks, e.g. at end of proof
\RequirePackage{graphicx}
\RequirePackage{amsmath}
\RequirePackage{amssymb}
\RequirePackage{amsfonts}
\usepackage{float}
\RequirePackage{units}
\newcommand{\pt}{p_{\perp}}
\journalname{Eur. Phys. J. C}
\begin{document}

\title{{\hfill \normalsize \rm{CERN-TH-2016-183, MCnet-16-33}\\ \medskip \\}Simulating V+jet processes in heavy ion collisions with JEWEL}

\author{Raghav Kunnawalkam Elayavalli\thanksref{e1,addr1, addr4}
        \and
        Korinna Christine Zapp\thanksref{e2,addr2,addr3,addr4} %etc.
}

\thankstext{e1}{e-mail: raghav.k.e@cern.ch}
\thankstext{e2}{e-mail: korinna.zapp@cern.ch}


\institute{Department of Physics and Astronomy, Rutgers University, Piscataway, New Jersey 08854, USA \label{addr1}
	\and CENTRA, Instituto Superior T\'ecnico, Universidade de Lisboa, Av. Rovisco Pais, P-1049-001 Lisboa, Portugal \label{addr2}
      \and
    Laborat\'{o}rio de Instrumenta\c{c}\~{a}o e F\'{\i}sica Experimental de Part\'{\i}culas (LIP), Av. Elias Garcia 14-1, 
1000-149 Lisboa, Portugal\label{addr3}
           \and 
           Theoretical Physics Department, CERN, CH-1211 Gen\`eve 23, Switzerland \label{addr4}
}

\date{Received: date / Accepted: date}
% The correct dates will be entered by the editorutilize

\maketitle

\begin{abstract}

Processes in which a jet recoils against an electroweak boson complement studies of jet quenching in heavy ion collisions at the LHC. As the boson does not interact strongly it escapes the dense medium unmodified and thus provides a more direct access to the hard scattering kinematics than can be obtained in di-jet events. First measurements of jet modification in these processes are now available from the LHC experiments and will improve greatly with better statistics in the future. We present an extension of \textsc{Jewel} to boson-jet processes. \textsc{Jewel} is a dynamical framework for jet evolution in a dense background based on perturbative QCD, that is in agreement with a large variety of jet observables. We also obtain a good description of the CMS and ATLAS data for $\gamma$+jet and $Z$+jet processes at \unit[2.76]{TeV} and \unit[5.02]{TeV}.

\keywords{Quark Gluon Plasma \and Jet quenching \and Vector bosons \and Transverse momentum imbalance}
% \subclass{MSC code1 \and MSC code2 \and more}utilize
\end{abstract}

\section{Introduction}

During Run I of the LHC, the modifications of jets due to re-scattering in the dense medium created in heavy ion collisions have been studied mostly in single-inclusive jet observables and di-jet events. They are dominated by pure QCD production processes, which have by far the largest cross sections. However, in these events it is practically impossible to determine the hard scattering kinematics, as all jets undergo quenching in the medium. This is different in $V$+jet processes, where a hard jet recoils against an electroweak gauge boson. The bosons -- and in the cases of $Z$ and $W$ production the leptonic decay products -- do not interact strongly and thus escape unmodified from the medium. This has been confirmed by measurements of inclusive vector boson production in Pb+Pb collisions at the LHC~\cite{Aad:2015lcb,Aad:2012ew,Aad:2014bha,Chatrchyan:2012vq,Chatrchyan:2014csa,Chatrchyan:2012nt}, which show that the observed rates are consistent with binary scaling and nuclear PDFs. The boson thus allows us to experimentally access the hard scattering kinematics. However, due to QCD corrections, in particular initial state radiation, the boson's and the parton's transverse momentum do not match exactly and the $\pt$ ratio fluctuates considerably from one event to another (cf. Fig.~\ref{fig:ypjet_xjy}). Nevertheless, since the initial parton $\pt$ is known on average, boson-jet processes provide valuable information that is complementary to pure QCD processes. First measurements~\cite{Chatrchyan:2012gt,atlasypjet,atlaszpjet,cmsypjet,cmszpjet} are still limited by statistics, but this will improve in future LHC running. There have also been attempts to study $\gamma$-hadron correlations at RHIC~\cite{Nguyen:2010wb,Hamed:2008yz}, but these are much more sensitive to poorly constrained hadronisation effects as opposed to jets.

The theoretical description of jet quenching in boson-jet events is the same as in pure jet events, in some approaches boson-jet~\cite{Wang:1996pe,Dai:2012am,Qin:2012gp,Ma:2013bia,Wang:2013cia,Casalderrey-Solana:2015vaa,Chien:2015hda,Chang:2016gjp} or $\gamma$-hadron~\cite{Wang:2010yz,Renk:2009ur,Zhang:2009fg} observables have been discussed specifically. Jet quenching calculations still struggle to describe all jet quenching observables at the same time. Boson+jet processes provide an important test for the predictions of jet quenching frameworks, that have already been constrained on other jet quenching data.

We here present an extension of \textsc{Jewel} to boson-jet processes\footnote{The code is available at \texttt{http://jewel.hepforge.org}.}. After a summary of the new features, we compare \textsc{Jewel} to boson-jet data from LHC Run I and II.

\section{Simulating $V$+jet processes with JEWEL}
\label{sec:jewel}

\textsc{Jewel} is a fully dynamical perturbative framework for jet quenching. It describes the simultaneous scale evolution of hard partons giving rise to jets and re-scattering in the medium. The former is implemented in the form of a virtuality ordered parton shower. All partons in the shower in addition to the jet evolution, undergo re-scattering in the background. These interactions are described by $2\to 2$ perturbative QCD matrix elements supplemented with parton showers and can thus be elastic or inelastic, where the two types of interactions occur with the (leading log) correct relative rates. This is the standard way of treating scattering processes in perturbative QCD and has a known and controlled formal accuracy (LO+LL in this implementation). However, its use in the context of re-scattering in a QGP in the \textsc{Jewel} framework goes beyond factorisation theorems and relies on a few assumptions, namely that (i) the re-scattering resolves the partonic structure of the QGP (which is certainly true for sufficiently hard interactions), (ii) an infra-red continuation can be invoked to regularise the pQCD matrix elements and include the dominant effect of soft scattering, (iii) the interplay of different sources of radiation is governed by the formation times and (iv) the physical picture of the LPM interference obtained in eikonal kinematics is also valid in the non-eikonal regime. For a full discussion of the \textsc{Jewel} framework and its implementation the reader is referred to~\cite{Zapp:2012ak}, here only the most important features will be summarised. 

The emissions due to the scale evolution of the jet get dynamically interleaved with radiation associated to re-scatterings in such a way that re-scattering can only induce radiation if its formation time is shorter than the lifetime of the hard parton. This implies that only a hard re-scattering can perturb hard parton shower emissions related to the initial jet production process, so that the hard jet structure is protected from medium modifications. This principle shares important features with colour coherence (cf. e.g.~\cite{CasalderreySolana:2012ef}), but is not a dynamical implementation of colour coherence. It is missing, for instance, soft and large angle emissions from coherent sub-systems. 

In \textsc{Jewel} the medium is fully dynamical and recoils in jet-medium interactions, thus giving rise to elastic energy loss (which also occurs in inelastic re-scatterings). The knowledge about the energy-momentum transferred from the jet to the medium can be used for detailed studies of the medium response to jets~\cite{Floerchinger:2014yqa}. \textsc{Jewel} has the option to retain recoiling medium partons in the event, but this requires special analysis techniques~\cite{bckgrnd}. For inclusive jet observables like the jet $\pt$ this leads to only small corrections, but certain jet-substructure observables are sensitive to the medium response. The observables discussed in this publication require only the jet $\pt$ and axis and are thus calculated without medium response.

All scattering processes within the formation time of a medium-induced emission act coherently, which means that only the vectorial sum of the momentum transfers matters for the gluon emission. This is the QCD analogue of the Landau-Pomerantchuk-Migdal effect, which is also implemented according to a generalisation of the algorithm derived in~\cite{Zapp:2011ya}. 

For jet evolution in vacuum \textsc{Jewel} reduces to a standard virtuality ordered final state parton shower. Initial state parton showers, hard jet production matrix elements, hadronisation and hadron decays are generated by \textsc{Pythia}\,6.4~\cite{Sjostrand:2006za}. The strong coupling $\alpha_s$ runs at one loop evaluated according to the standard perturbative scale choices. $\Lambda_\text{QCD}$ is adjusted to fit LEP data and is the same throughout the simulation.

\smallskip

\begin{figure}[h!] %  figure placement: here, top, bottom, or page
   \centering
   \includegraphics[width=0.45\textwidth]{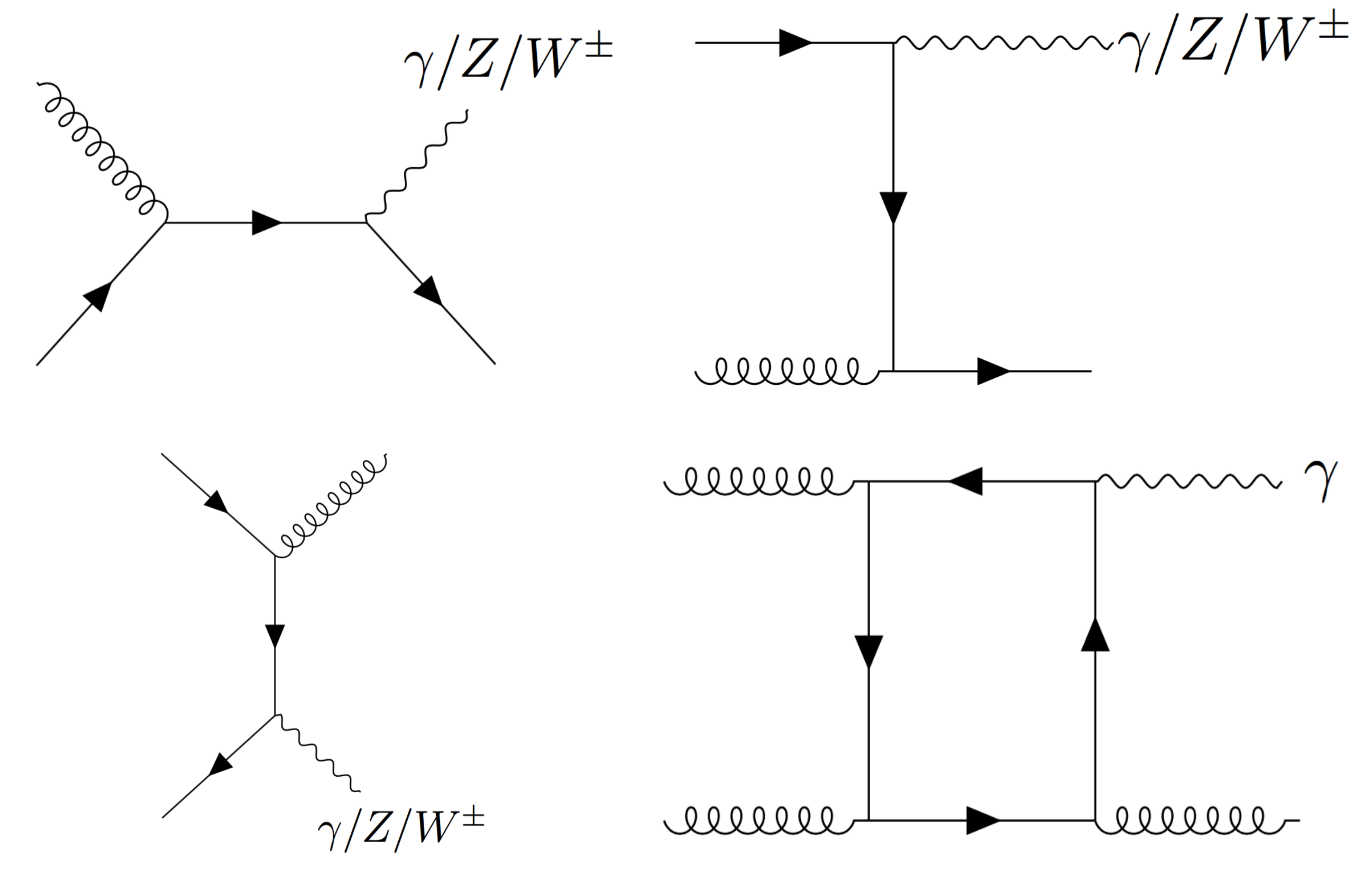} 
   \caption{Feynman diagrams for $V$+jet processes included in \textsc{Jewel}}
   \label{fig:production}
\end{figure}

In the extended version we have included the lowest order processes with a jet recoiling against a
vector boson. The corresponding  diagrams are shown in
Fig.~\ref{fig:production}. These correspond to either a quark scattering off a
gluon (Compton scattering) or a quark--anti-quark pair annihilating to produce a
boson and a gluon. For photons, the box diagram $gg\to \gamma g$ is also
included. This process is of higher order than the others, but is included as it can be numerically important in certain phase space regions.
The leptonic decays of the heavy boson $Z$ and $W$ are simulated as
well. 

Hard photons can also be radiated off quarks during jet evolution. These
fragmentation photons are typically accompanied by hadronic activity and are 
suppressed by requiring the photon to be isolated. However, it is still
possible that fragmentation photons pass the isolation criterion. The probability
for this to happen is small and depends on the cuts. It has to the
best of our knowledge not been quantified in a heavy ion environment in the
presence of jet quenching. In the current \textsc{Jewel} version 
fragmentation photons are also not included. For the analyses shown here the fragmentation component is expected to be small due to the applied photon isolation.

\textsc{Jewel} is a leading-order framework. While NLO corrections to $V$+jet processes can be sizable, in the observables shown here corrections affecting only the cross section largely cancel due to the normalisation to number of bosons or number of boson-jet pairs. The corrections to differential distributions remain, but are typically smaller.

\subsection{The new parameters and Switches}

We have expanded the parameter set listed in~\cite{Zapp:2013vla} as follows
(default values are given in parentheses).

\begin{description}
\item[PROCESS (`PPJJ'):] process that is to be simulated by matrix element,
available options are

\begin{description}
 \item[`EEJJ':] di-jet production in $e^+$+$e^-$ collisions
 \item[`PPJJ':] di-jet production in hadronic collisions
 \item[`PPYJ':] all $\gamma$+jet processes
 \item[`PPYQ':] only $\gamma$+quark production
 \item[`PPYG':] only $\gamma$+gluon production
 \item[`PPZJ':] all $Z$+jet processes
 \item[`PPZQ':] only $Z$+quark production
 \item[`PPZG':] only $Z$+gluon production
 \item[`PPWJ':] all $W^\pm$+jet processes
 \item[`PPWQ':] only $W^\pm$+quark production
 \item[`PPWG':] only $W^\pm$+gluon production
\end{description}

\item[CHANNEL (`MUON'):] decay channel for the heavy $W$ and $Z$ bosons,
available are `ELEC' and `MUON' for the decay to electrons/positrons and muons,
respectively

\item[ISOCHANNEL ('XX'):] isospin channel for the hard matrix element, can be 'PP', 'PN', 'NP' or 'NN' to select the proton-proton, proton-neutron, neutron-proton or neutron-neutron channel, respectively. For all other values all four channels will be simulated with the correct relative weights.

\item[NPROTON (82):] number of protons in the nucleus
\end{description}

\section{Comparisons to data}

We generate events in the standard setup~\cite{Zapp:2013vla} at $\sqrt{s_\text{NN}} = \unit[2.76]{TeV}$ and $\sqrt{s_\text{NN}} = \unit[5.02]{TeV}$ with the simple parametrisation of the background discussed in detail in~\cite{Zapp:2013zya}. This background model describes a thermal quark-gluon gas undergoing Bjorken expansion with a superimposed transverse profile obtained from an optical Glauber model. 
The initial conditions for the background model are initial time $\tau_\text{i}=\unit[0.6]{fm}$ and temperature $T_\text{i}=\unit[485]{MeV}$ for $\sqrt{s_\text{NN}} = \unit[2.76]{TeV}$~\cite{Shen:2012vn} and $\tau_\text{i}=\unit[0.4]{fm}$ and $T_\text{i}=\unit[590]{MeV}$ for $\sqrt{s_\text{NN}} = \unit[5.02]{TeV}$~\cite{Shen:2014vra}. They are taken from a hydrodynamic calculation describing soft particle production.
The proton PDF set is \textsc{Cteq6LL}~\cite{Pumplin:2002vw} and for the Pb+Pb sample the \textsc{Eps09}~\cite{Eskola:2009uj} nuclear PDF set is used in addition, both are provided by \textsc{Lhapdf}~\cite{Whalley:2005nh}. The only parameter in \textsc{Jewel} that can be fitted to jet quenching data is the scaling factor of the Debye mass. It was adjusted once to describe the single-inclusive hadron suppression at RHIC and has remained the same since.

We use the \textsc{Rivet} analysis framework~\cite{Buckley:2010ar} for all our
studies. Jets are reconstructed using the same jet algorithm as the experiments
(anti-$k_\perp$~\cite{Cacciari:2008gp}) from the \textsc{FastJet}
package~\cite{Cacciari:2011ma}.

\subsection{$\gamma$+jet}

\begin{figure*}
 \includegraphics[width=0.5\textwidth]{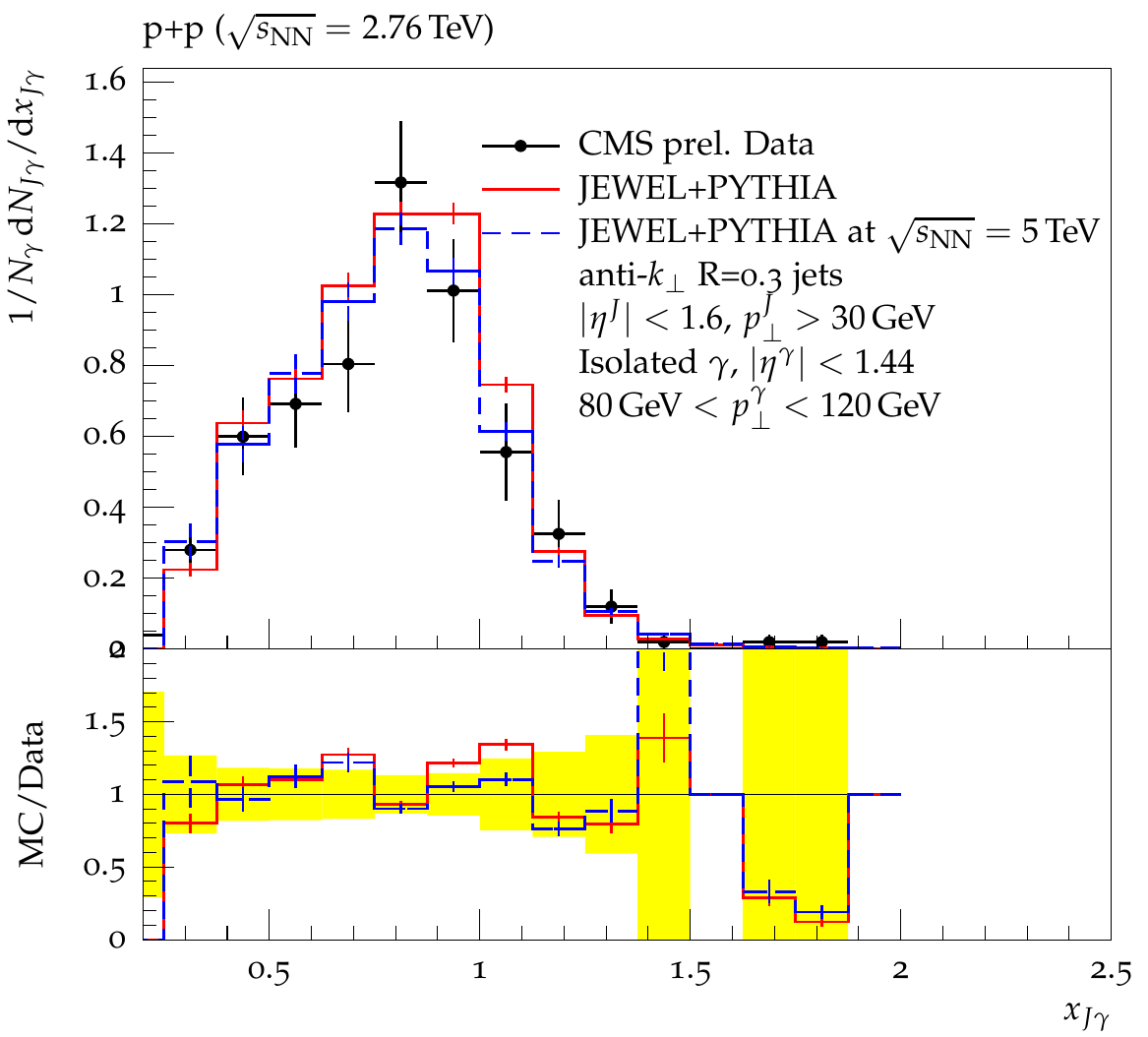}
 \includegraphics[width=0.5\textwidth]{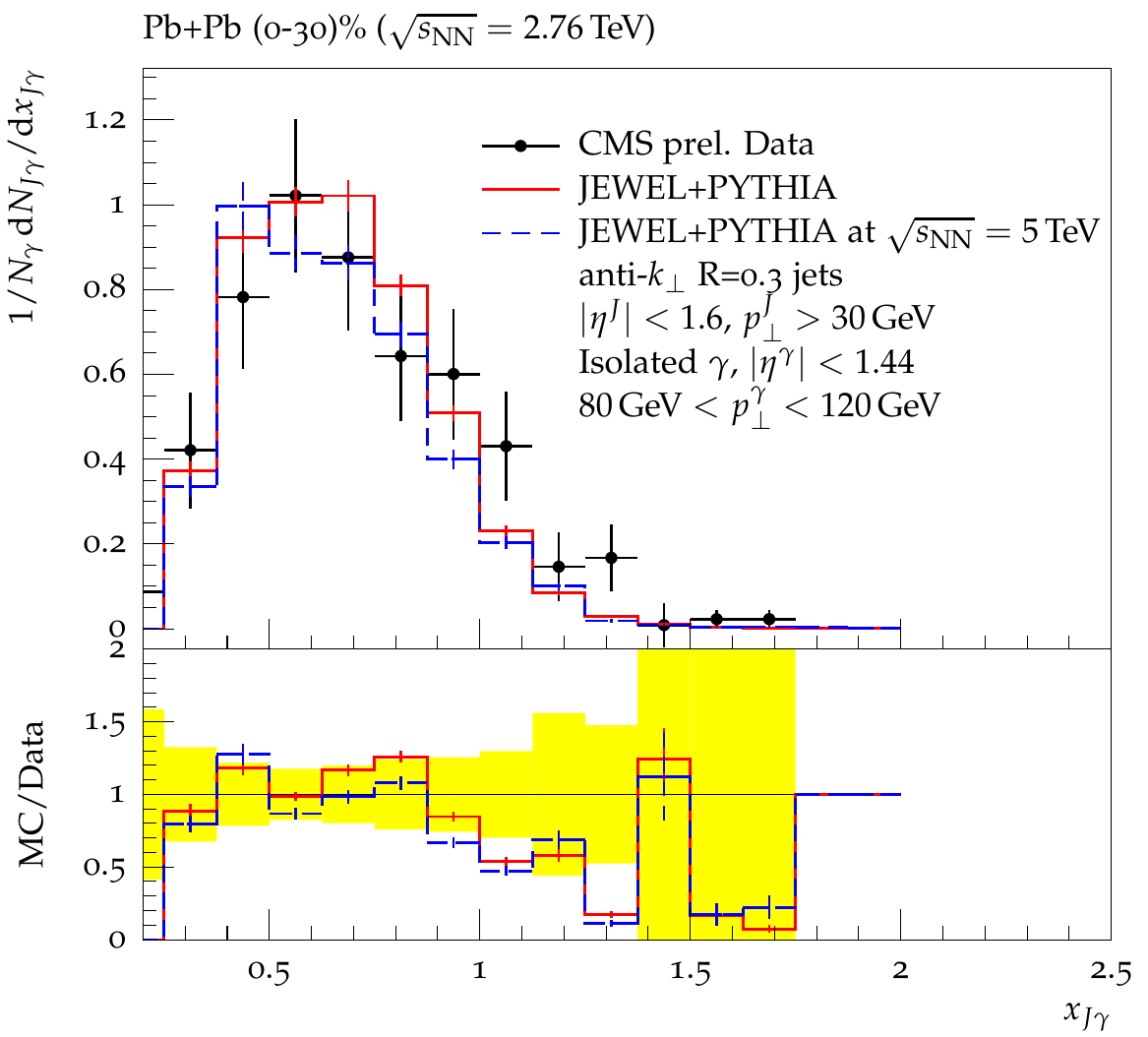}
\caption{Momentum imbalance $x_{J \gamma} = \pt^J/\pt^\gamma$ in $\gamma$+jet events for photon transverse momentum $\unit[80]{GeV} < \pt^\gamma < \unit[120]{GeV}$ compared to preliminary CMS data~\cite{cmsypjet} in p+p (left) and central Pb+Pb events (right) at $\sqrt{s_\text{NN}} = \unit[2.76]{TeV}$. The \textsc{Jewel+Pythia} prediction for  $\sqrt{s_\text{NN}} = \unit[5.02]{TeV}$ is also shown. The CMS data are not unfolded for jet energy resolution, therefore the jet $\pt$ was smeared in the Monte Carlo sample using the parametrisation from~\cite{Chatrchyan:2012gt}. 
The data points have been read off the plots and error bars correspond to statistical errors only. The yellow band in the ratio plot indicates the errors on the data points.}
\label{fig:ypjet_xjy}
\end{figure*}

As discussed in Section~\ref{sec:jewel}, the background from fragmentation and
decay photons has to be suppressed. Therefore the photon is demanded to be
isolated by requiring the sum of energy in a cone of radius $0.4$ (in the
$\eta-\phi$ phase space where $\eta$ is the pseudorapidity and $\phi$ is the azimuthal angle in the plane transverse to the beam axis) around the photon to be less than $\unit[7]{\%}$ of the
photon's energy. In addition, the photon has to be within $|\eta^\gamma| < 1.44$
and have a transverse momentum $\pt^\gamma > \unit[40]{GeV}$. The jets are
reconstructed with the anti-$k_\perp$ algorithm with a resolution parameter of
$R=0.3$. Jets are required to have a $\pt^J>\unit[30]{GeV/c}$ and to be in the
barrel region ($|\eta^J|<1.6$). Furthermore, only jets that are back-to-back with
the photon ($\Delta \phi_{J\gamma} > 7\pi/8$) are selected.

\begin{figure*}
 \includegraphics[width=0.5\textwidth]{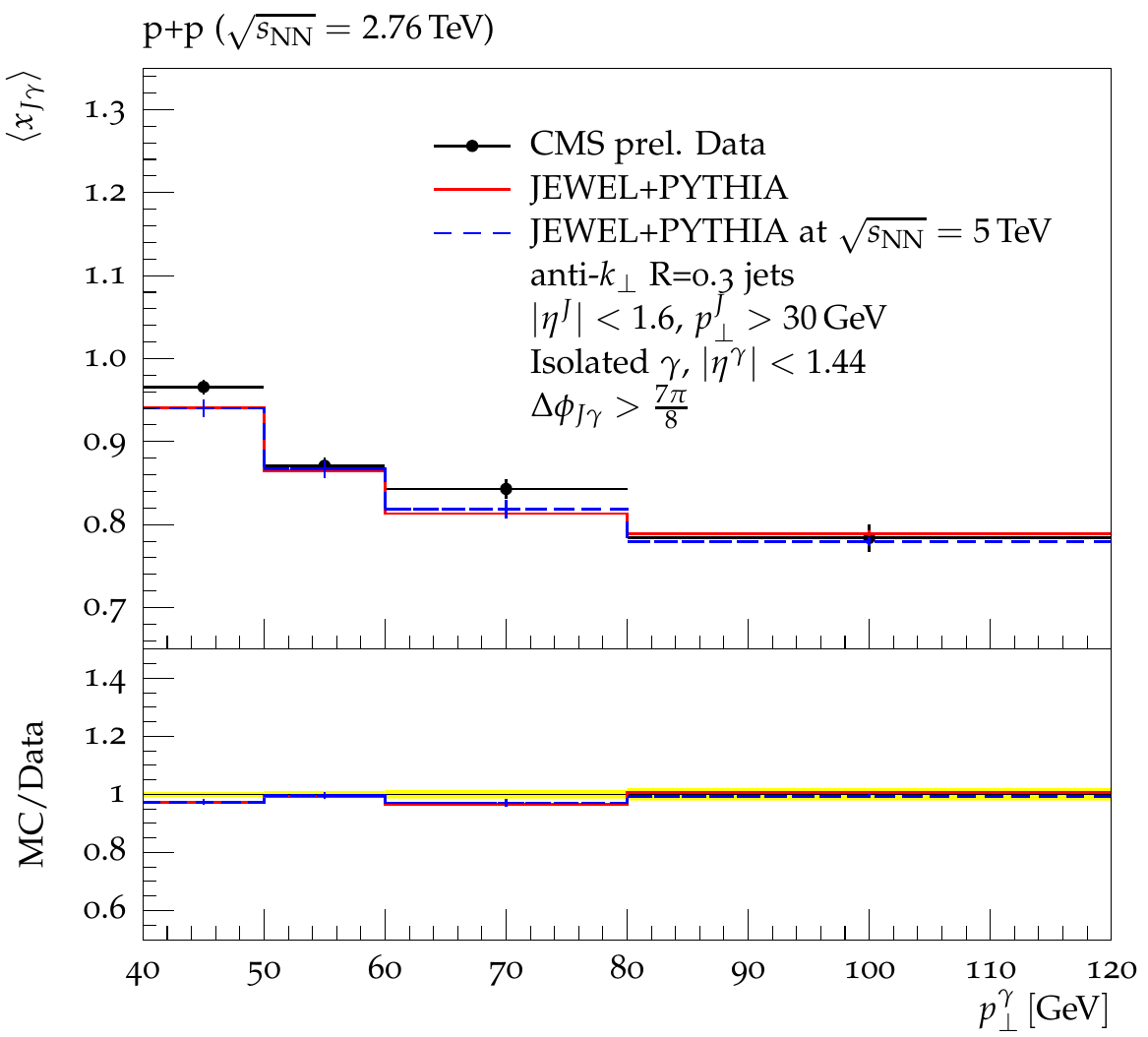}
 \includegraphics[width=0.5\textwidth]{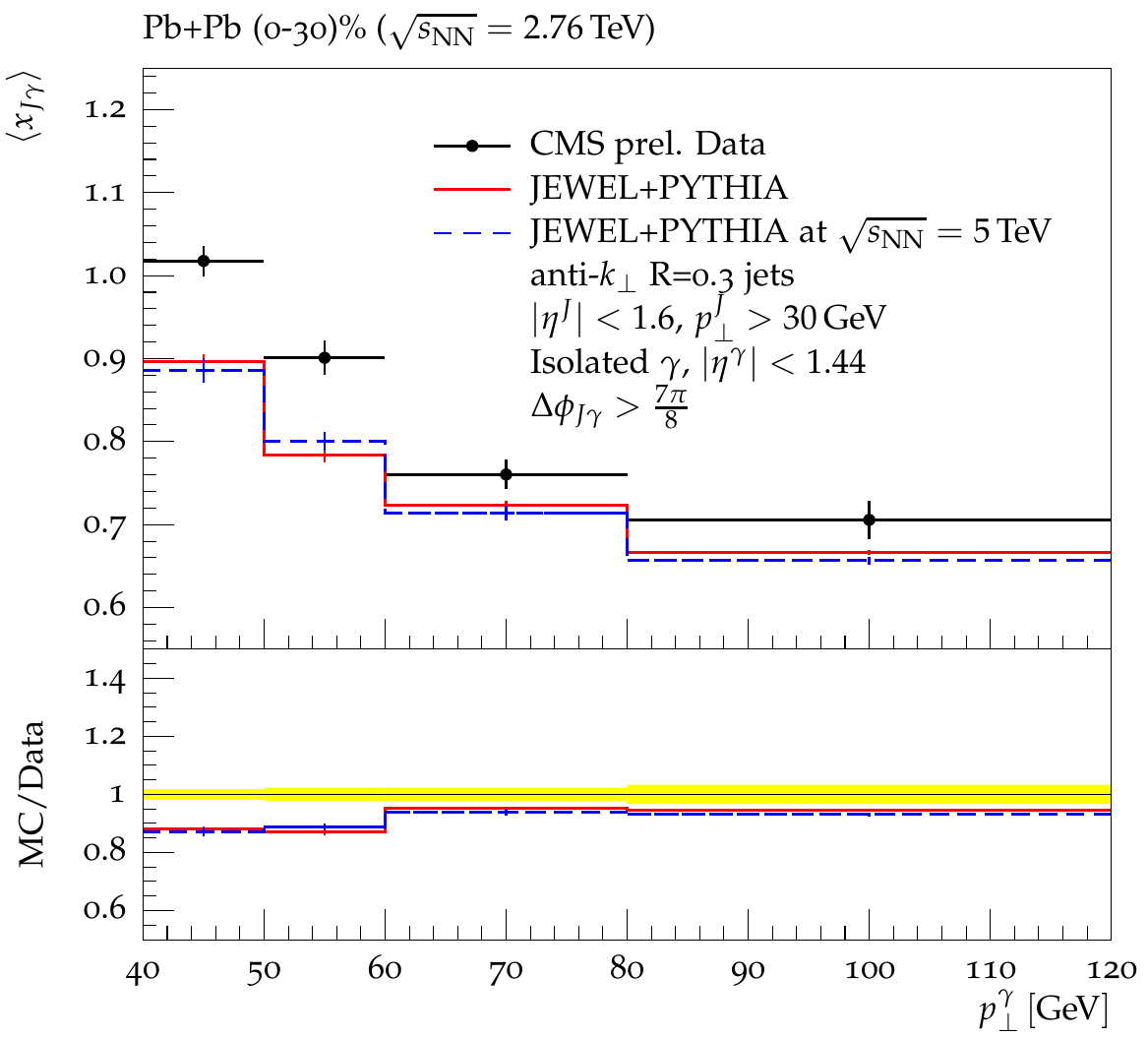}
\caption{Average value of the $x_{J \gamma}$ shown as a
function of the
photon's transverse momentum compared to preliminary CMS data~\cite{cmsypjet} in p+p (left) and central Pb+Pb events (right) at $\sqrt{s_\text{NN}} = \unit[2.76]{TeV}$. The \textsc{Jewel+Pythia} prediction for  $\sqrt{s_\text{NN}} = \unit[5.02]{TeV}$ is also shown. The CMS data are not unfolded for jet energy resolution, therefore the jet $\pt$ was smeared in the Monte Carlo sample using the parametrisation from~\cite{Chatrchyan:2012gt}. 
The data points have been read off the plots and error bars correspond to statistical errors only. The yellow band in the ratio plot indicates the errors on the data points.}
\label{fig:ypjet_average}
\end{figure*}

\begin{figure*}
 \includegraphics[width=0.5\textwidth]{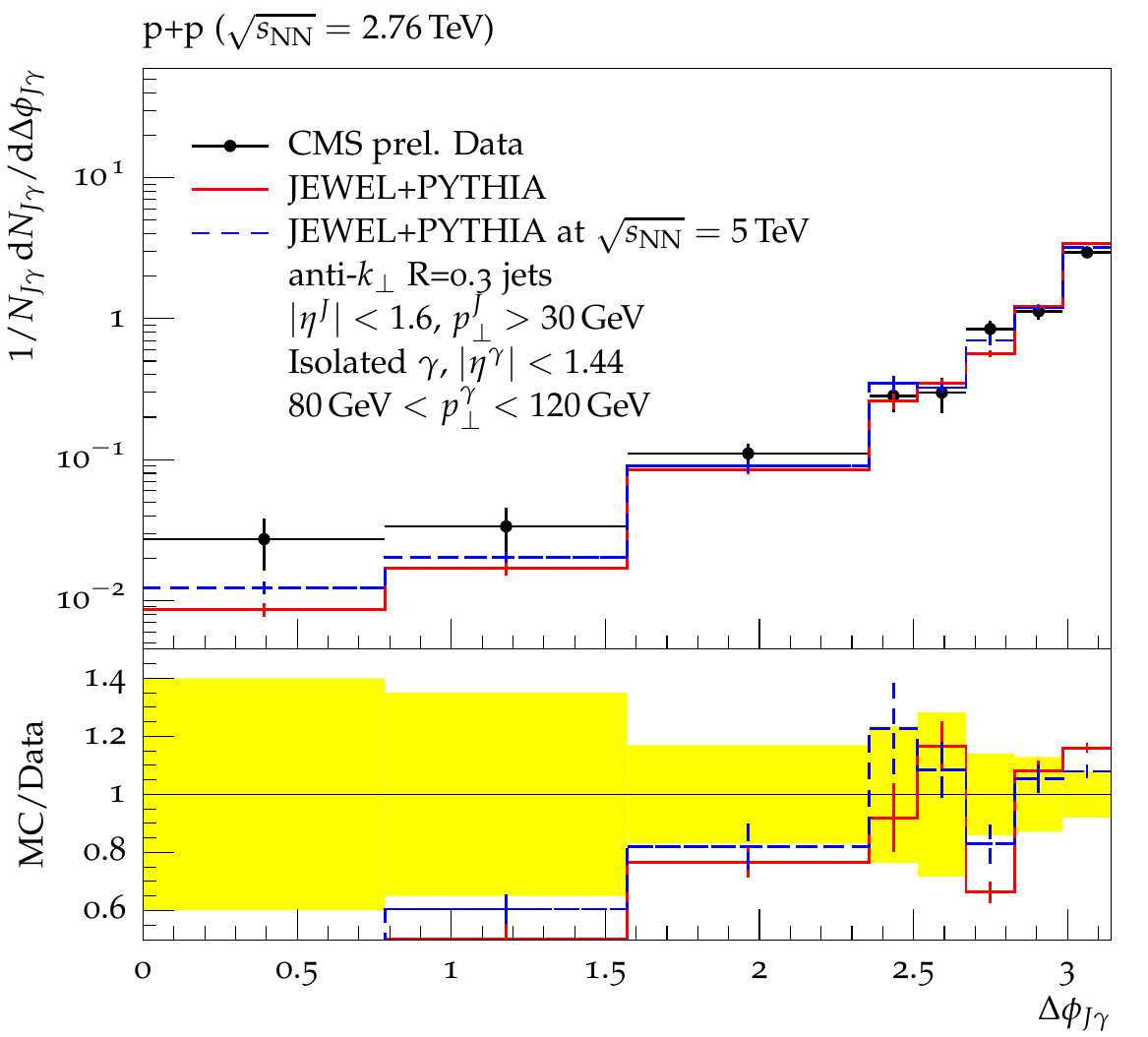}
 \includegraphics[width=0.5\textwidth]{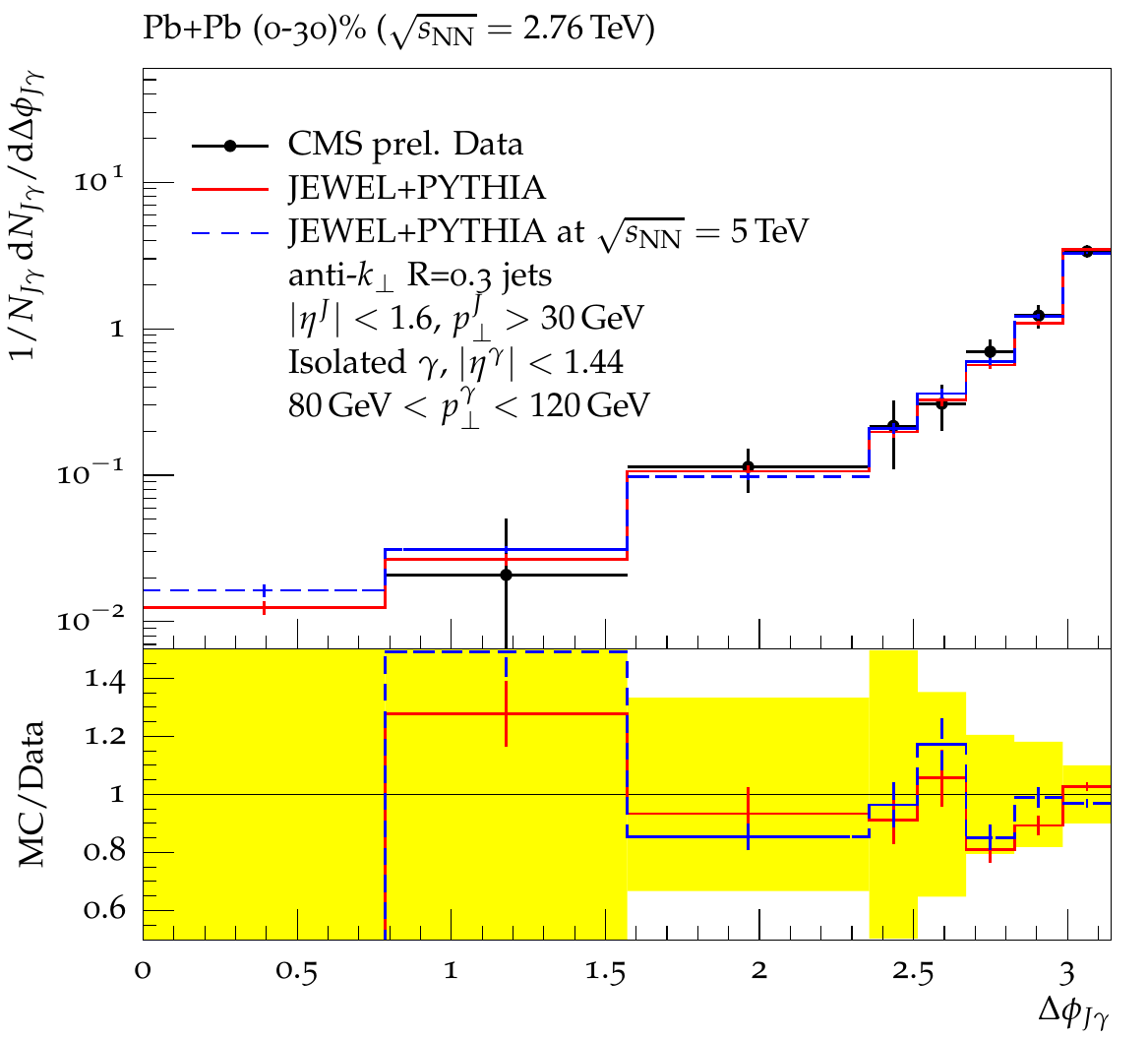}
\caption{Azimuthal angle $\Delta \phi_{J\gamma}$ between the photon and the jet for photon transverse momentum $\unit[80]{GeV} < \pt^\gamma < \unit[120]{GeV}$ compared to preliminary CMS data~\cite{cmsypjet} in p+p (left) and central Pb+Pb events (right) at $\sqrt{s_\text{NN}} = \unit[2.76]{TeV}$. The \textsc{Jewel+Pythia} prediction for  $\sqrt{s_\text{NN}} = \unit[5.02]{TeV}$ is also shown. The CMS data are not unfolded for jet energy resolution, therefore the jet $\pt$ was smeared in the Monte Carlo sample using the parametrisation from~\cite{Chatrchyan:2012gt}. 
The data points have been read off the plots and error bars correspond to statistical errors only. The yellow band in the ratio plot indicates the errors on the data points.}
\label{fig:ypjet_deltaphi}
\end{figure*}

Fig.~\ref{fig:ypjet_xjy} shows our results for the transverse momentum asymmetry
in $\gamma+$jet pairs ($x_{J\gamma} = \pt^J/\pt^\gamma$) compared with preliminary CMS~\cite{cmsypjet} data
points for p+p and central ($0-30\%$) Pb+Pb collisions at $\sqrt{s_\text{NN}} = \unit[2.76]{TeV}$. 
Fig~\ref{fig:ypjet_average} shows the average value of the
$x_{J\gamma}$ as
a function of the photon transverse momenta in four $\pt$ bins, again for p+p and central
Pb+Pb collisions. \textsc{Jewel+Pythia} is able to
reproduce the effect of the $\pt$ imbalance for $\gamma+$jets events very nicely
for both p+p and Pb+Pb events. In central Pb+Pb collisions $\langle x_{J\gamma}\rangle$ is slightly lower in \textsc{Jewel+Pythia} than in the data indicating stronger medium modifications in \textsc{Jewel}, particularly at relatively low photon $\pt$. In Fig.~\ref{fig:ypjet_deltaphi} the azimuthal angle ($\Delta \phi_{J \gamma}$) between the photon and the jet is shown. We again find a very reasonable agreement with \textsc{Jewel+Pythia} for pp collisions slightly more peaked. In all three figures we also show the \textsc{Jewel+Pythia} predictions for $\sqrt{s_\text{NN}} = \unit[5.02]{TeV}$, which turn out to be very similar to the $\sqrt{s_\text{NN}} = \unit[2.76]{TeV}$ results. The agreement with the ATLAS
measurement~\cite{atlasypjet} is of a very similar quality. 

\subsection{$Z/W$+jet}

\begin{figure}
   \includegraphics[width=0.5\textwidth]{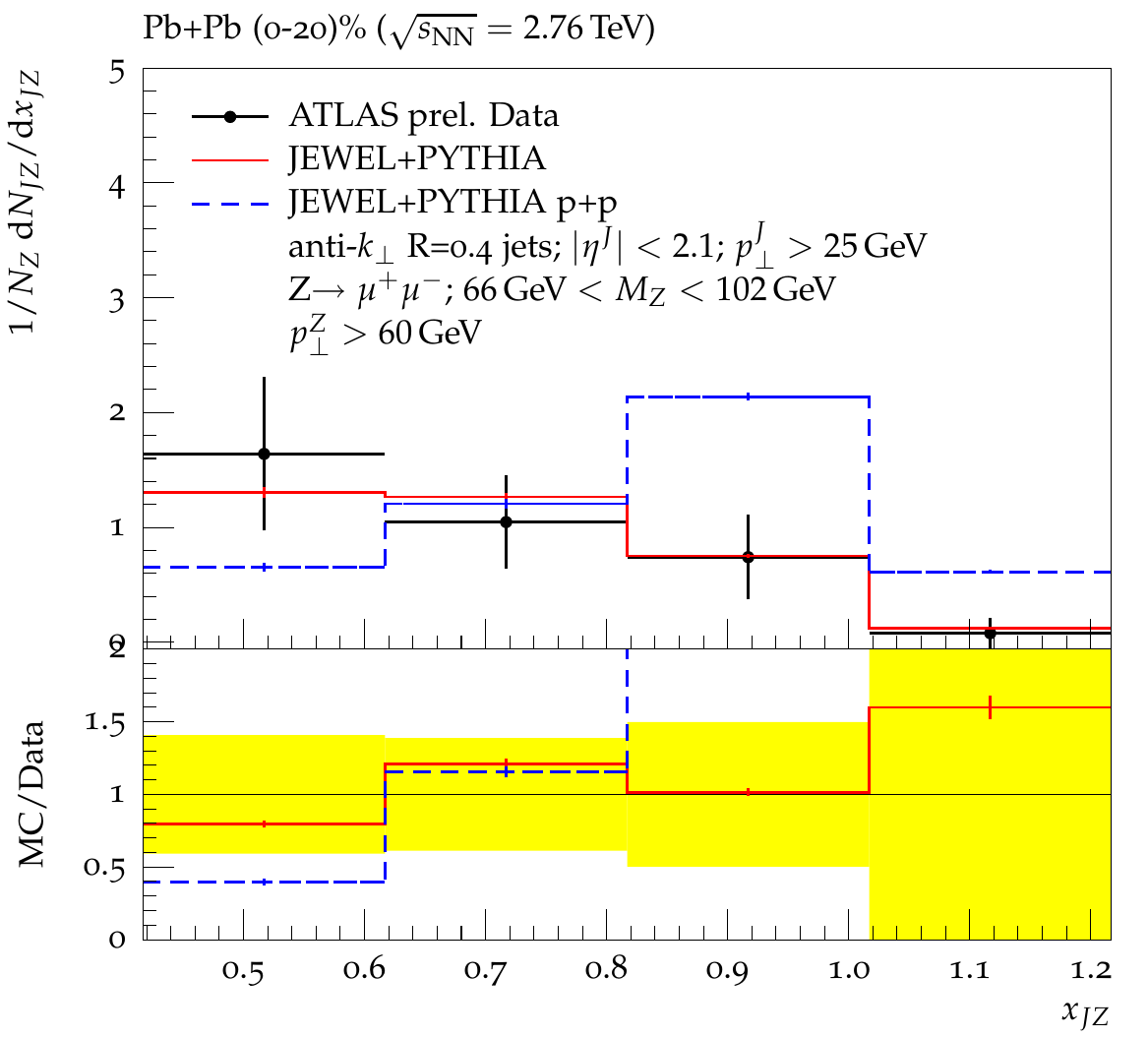} 
   \caption{Momentum imbalance $x_{JZ}$ in $Z$+jet events compared to preliminary ATLAS data~\cite{atlaszpjet} in central Pb+Pb events at $\sqrt{s_\text{NN}} = \unit[2.76]{TeV}$. The data points have been read off the plots and error bars correspond to statistical errors only. The yellow band in the ratio plot indicates the errors on the data points.}
   \label{fig:zpjet_atlas_r4}
\end{figure}

In the case of $Z$ and $W$ production we utilise the muon decay channel in our simulations (this is purely convenience, the electron channel can be simulated as well). The $Z$ candidate's momentum is reconstructed from the di-muon pairs. For comparison to the ATLAS measurement we require its reconstructed mass in the window $\unit[66]{GeV} < M_Z < \unit[102]{GeV}$ and $\pt^Z>\unit[60]{GeV}$. The jets are reconstructed with the same anti-$k_\perp$ algorithm with resolution parameter $R=0.4$, with the kinematic cut on its $\pt^J>\unit[25]{GeV}$ and it is required to be found in the barrel region $|\eta^J|<2.1$. Similar to the $\gamma+$jet case, we impose $\Delta \phi_{JZ} > \pi/2$ to select the back to back pairs. 
Fig.~\ref{fig:zpjet_atlas_r4} shows the ATLAS~\cite{atlaszpjet} preliminary result for the $\pt$ imbalance compared to \textsc{Jewel+Pythia} for central (0-20\%) Pb+Pb collisions at  $\sqrt{s_\text{NN}} = \unit[2.76]{TeV}$. For comparison we also show the \textsc{Jewel+Pythia} result for p+p. In central Pb+Pb events we observe a clear shift of the distribution towards smaller $x_{JZ}$ compared to p+p and a reasonable agreement between the MC and data.

\begin{figure*}
   \includegraphics[width=0.5\textwidth]{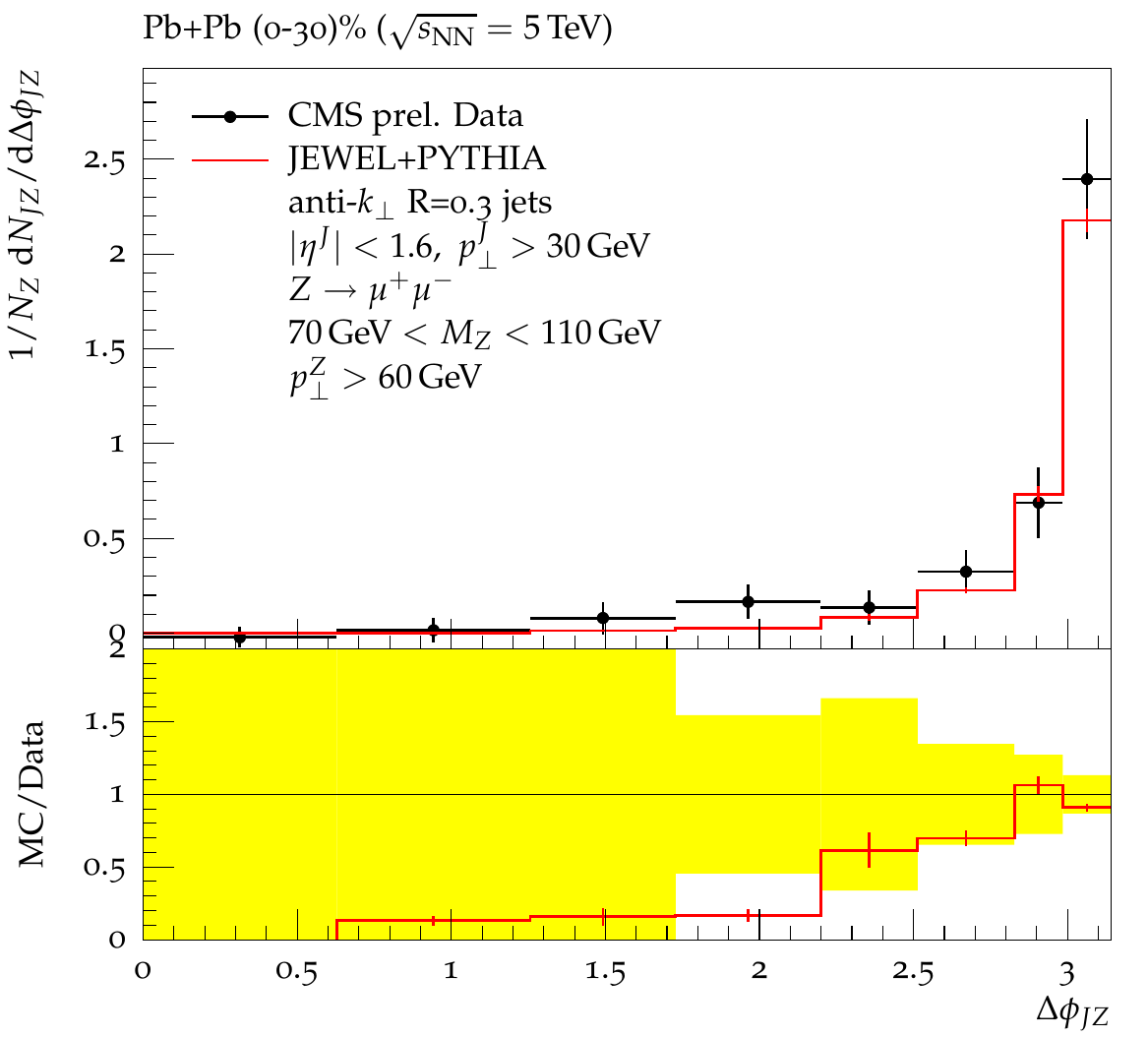} 
   \includegraphics[width=0.5\textwidth]{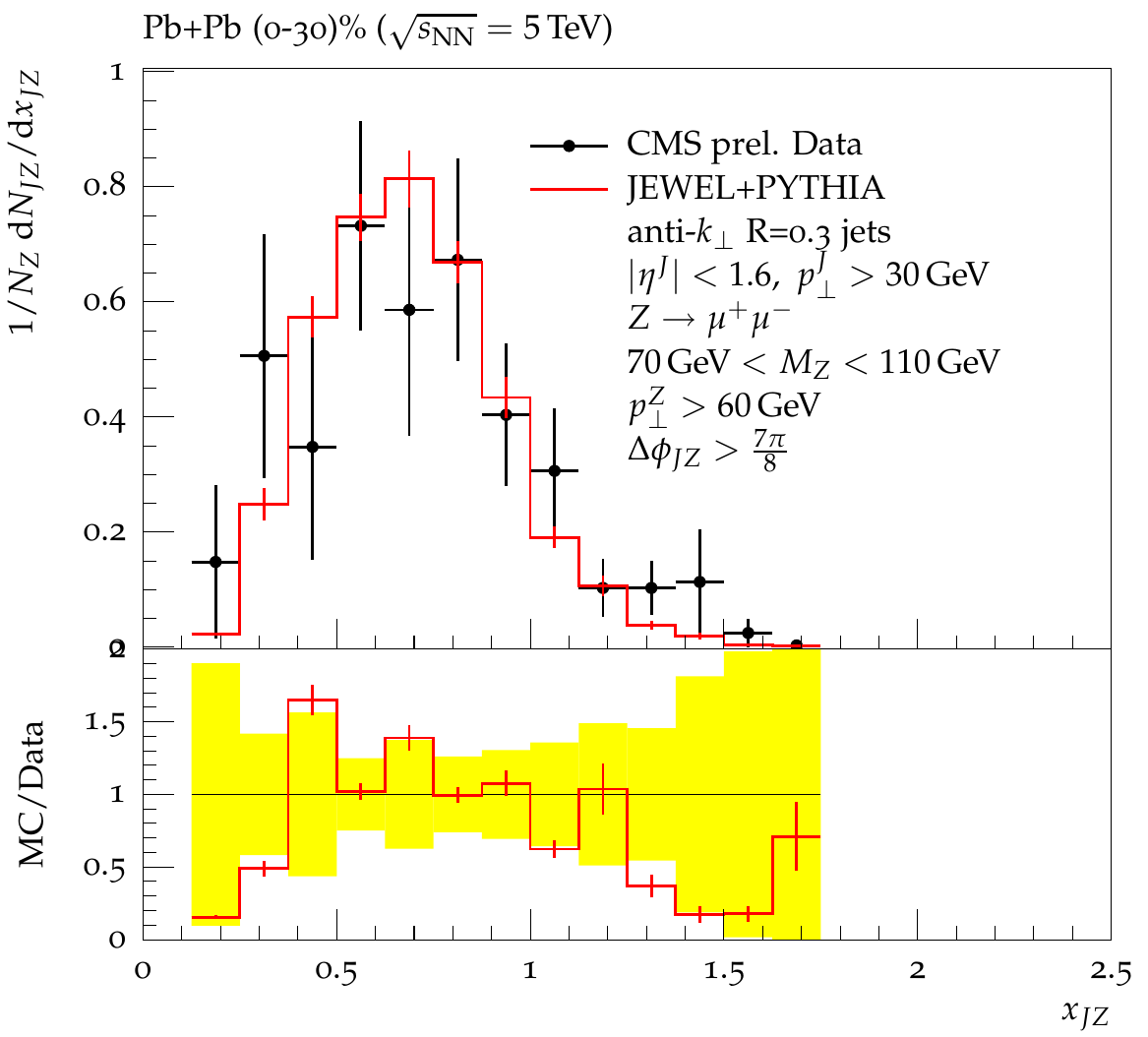} 
   \caption{Azimuthal angle $\Delta \phi_{JZ}$ between the $Z$ and the jet (left) and momentum imbalance $x_{JZ}$ (right) in $Z$+jet events compared to preliminary CMS data~\cite{cmszpjet} in central Pb+Pb events at $\sqrt{s_\text{NN}} = \unit[5.02]{TeV}$. The CMS data are not unfolded for jet energy resolution, therefore the jet $\pt$ was smeared in the Monte Carlo sample using the parametrisation from~\cite{Chatrchyan:2012gt}. The data points have been read off the plots and error bars correspond to statistical errors only. The yellow band in the ratio plot indicates the errors on the data points.}
   \label{fig:zpjet_cms_r3}
\end{figure*}

\begin{figure*}
   \includegraphics[width=0.5\textwidth]{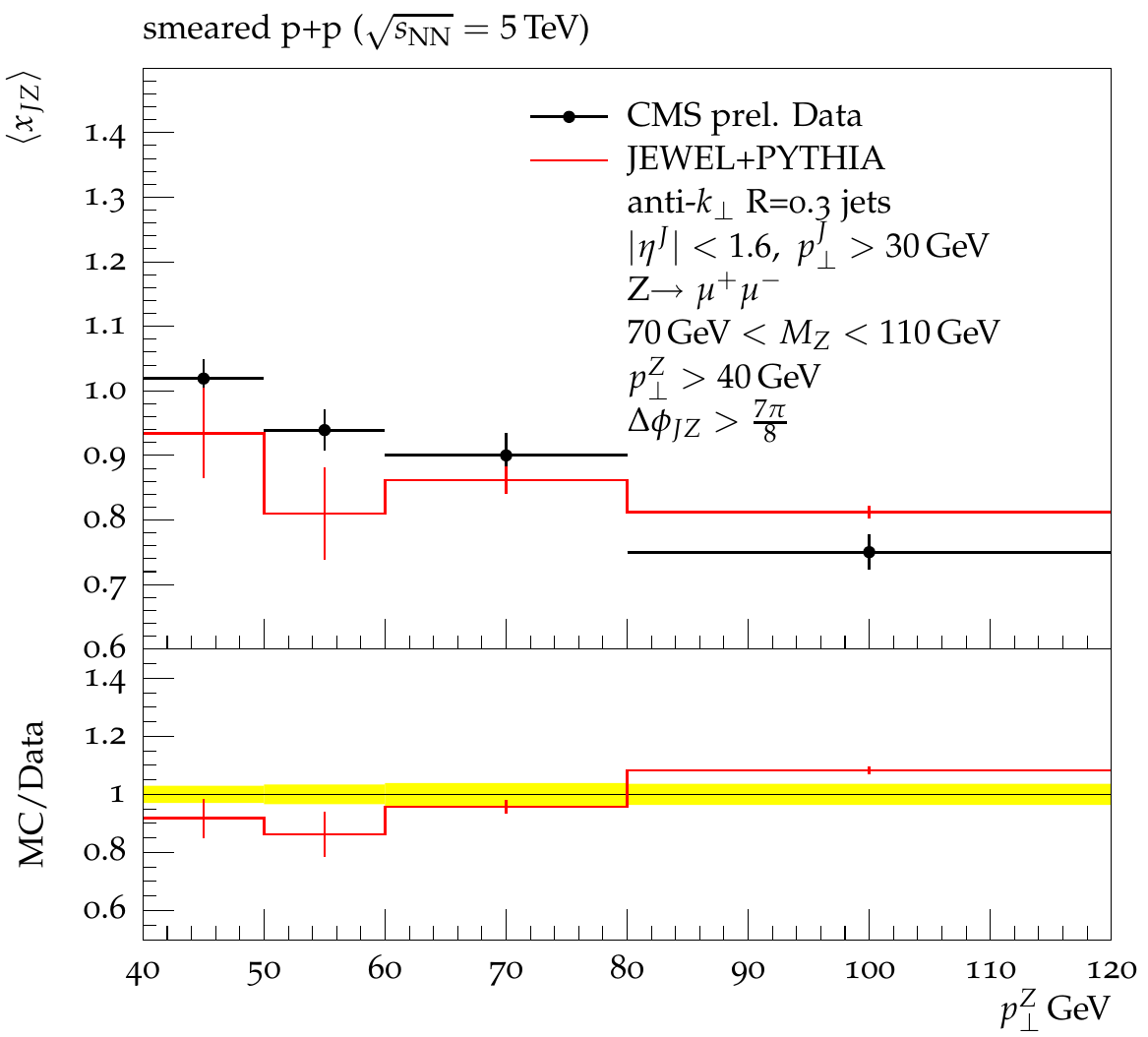} 
   \includegraphics[width=0.5\textwidth]{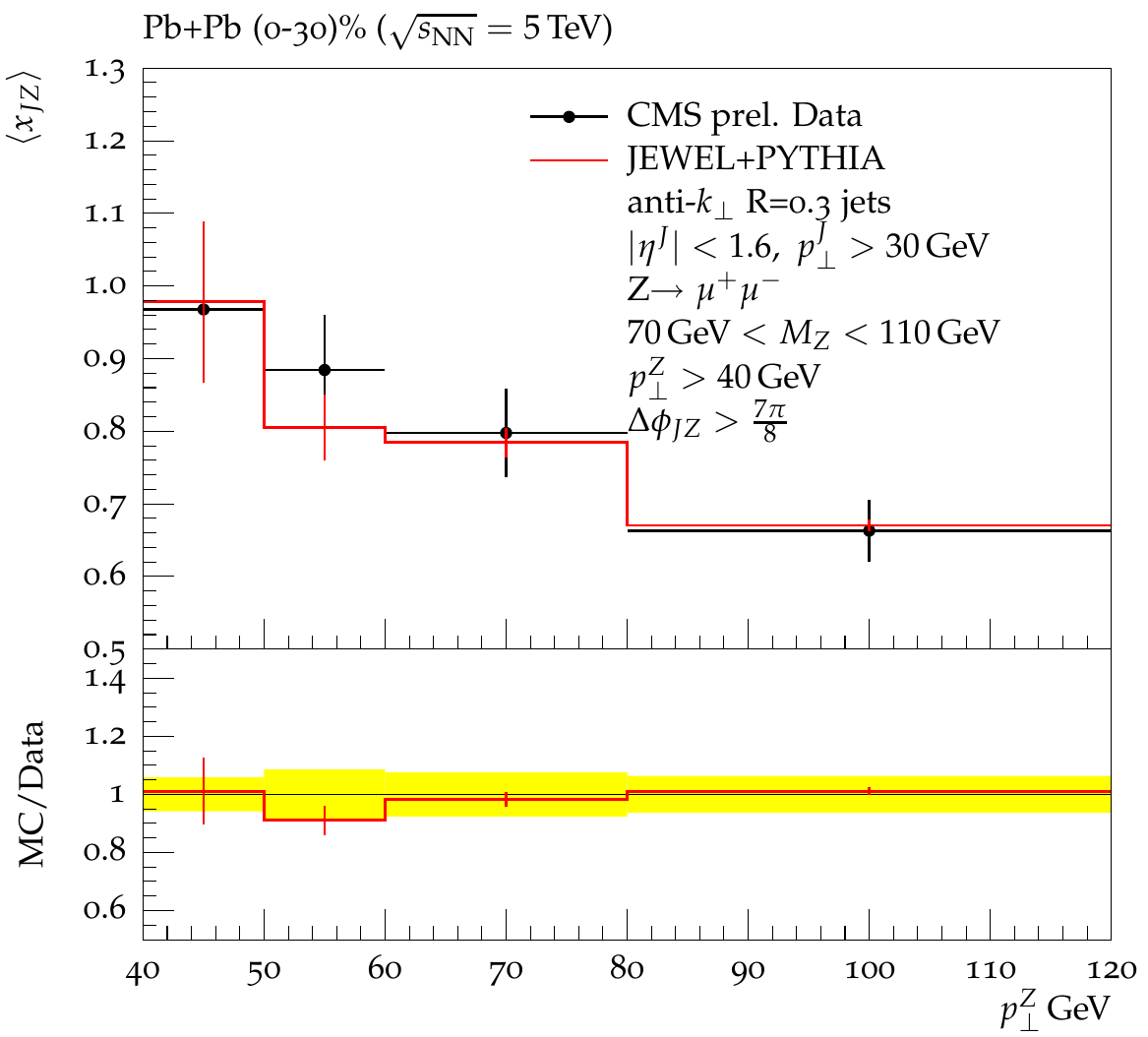} 
   \caption{Average value of the $x_{JZ}$ shown as a
function of the $Z$ transverse momentum compared to preliminary CMS data~\cite{cmszpjet} in p+p (left) and central Pb+Pb events (right) at $\sqrt{s_\text{NN}} = \unit[5.02]{TeV}$. The p+p sample has been smeared to match the resolution of the central Pb+Pb sample in data and \textsc{Jewel+Pythia}. The CMS data are not unfolded for jet energy resolution, therefore the jet $\pt$ was smeared in the Monte Carlo sample using the parametrisation from~\cite{Chatrchyan:2012gt}. 
The data points have been read off the plots and error bars correspond to statistical errors only. The yellow band in the ratio plot indicates the errors on the data points.}
   \label{fig:zpjet_cms_avcjZ}
\end{figure*}

In the CMS analysis jets are reconstructed with the anti-$k_\perp$ algorithm with resolution parameter $R=0.3$ and the cuts are  $\unit[70]{GeV} < M_Z < \unit[110]{GeV}$ and $\pt^Z>\unit[40]{GeV}$, $\pt^J>\unit[30]{GeV}$, $|\eta^J|<1.6$ and $\Delta \phi_{JZ} > 7\pi/8$. Fig.~\ref{fig:zpjet_cms_r3} shows the latest CMS $Z$+jet~\cite{cmszpjet} preliminary results for the azimuthal angle, $\Delta \phi_{J Z}$ between the jet and the $Z$ and the momentum imbalance $x_{JZ}$ at $\sqrt{s_\text{NN}}=\unit[5.02]{TeV}$ for central events ($0-30\%$). \textsc{Jewel+Pythia} nicely reproduces the  $x_{JZ}$ distribution, but once more the pairs are slightly more back-to-back than in data.

\begin{figure}
   \includegraphics[width=0.5\textwidth]{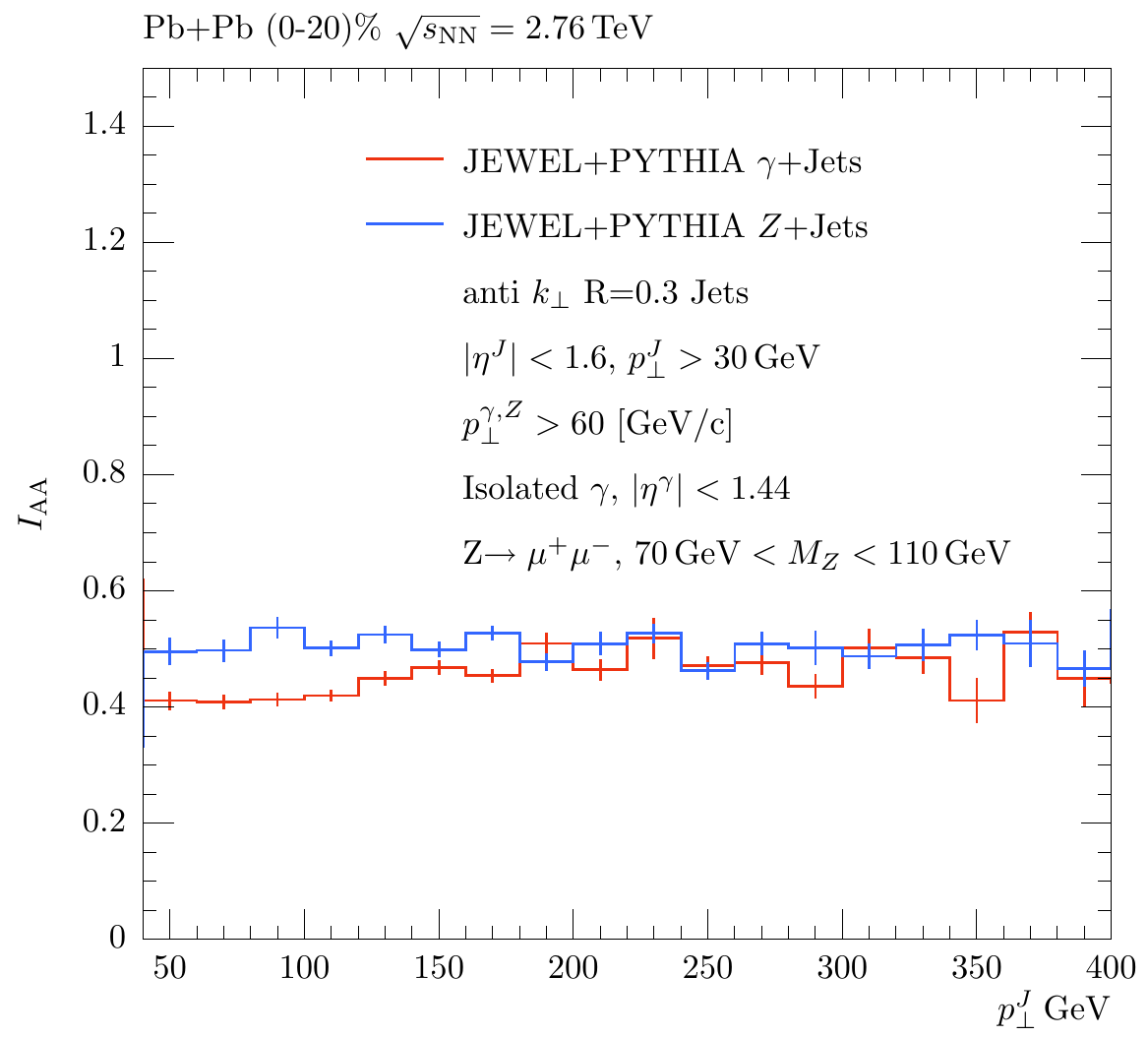} 
   \caption{Nuclear modification factor of the jet in $Z$+jet (blue) and $\gamma$+jet (red) events in central Pb+Pb events at $\sqrt{s_\text{NN}} = \unit[2.76]{TeV}$.}
   \label{fig:zyjet_raa}
\end{figure}

It is also informative to look at the nuclear modification factors ($I_{AA}$) of jets in events recoiling against a $\gamma$ or a $Z$. Due to the large mass of the $Z$ boson, the jet spectrum is harder than for jets recoiling off a $\gamma$. This influences the $I_{AA}$ for $Z$+jets to be less suppressed at the low $\pt$ range as shown in Fig.~\ref{fig:zyjet_raa}. 

Reconstructing a $W$ boson candidate in the heavy ion environment is difficult due to the ambiguous nature of the missing transverse energy (MET) in the event. Due to in-medium energy loss, the MET in such events does not accurately represent the neutrino, required to reconstruct the $W$. We therefore investigate the possibility of using the charged decay lepton instead of a reconstructed $W$. In both cases we require the lepton to have a high $\pt^\mu > \unit[60]{GeV}$ and $|\eta^\mu| < 2.5$, for reconstructed $W$'s the mass window is $\unit[60]{GeV} < M_W < \unit[100]{GeV}$. Jets are reconstructed with $R=0.4$ and kinematic cuts $\pt^J > \unit[25]{GeV}$ and $|\eta^J| < 2.1$. We also impost a $\Delta R_{J\mu}>0.6$ to ensure no overlap between our reconstructed jet and lepton collections. 

\begin{figure*}
   \includegraphics[width=0.5\textwidth]{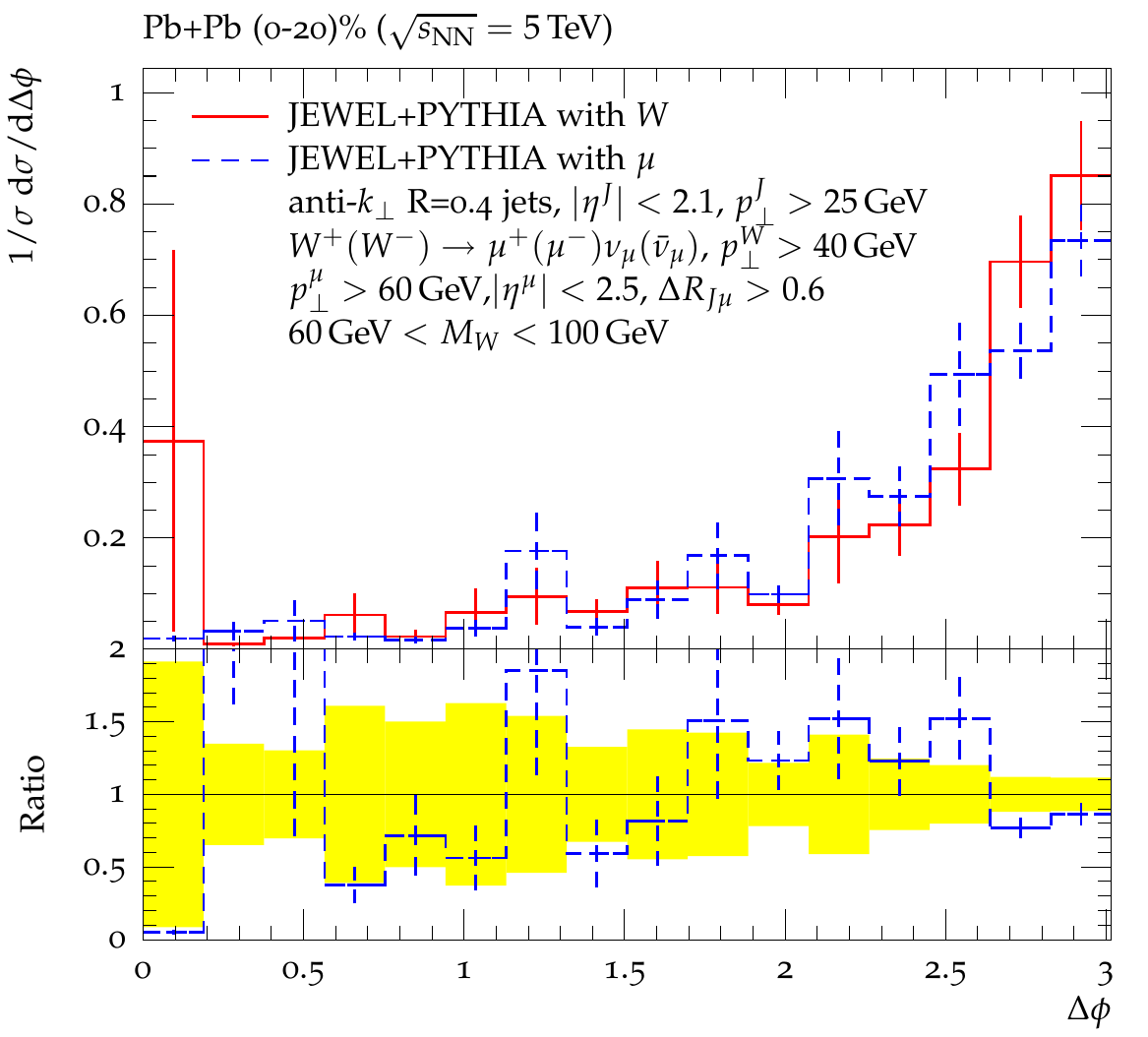} 
   \includegraphics[width=0.5\textwidth]{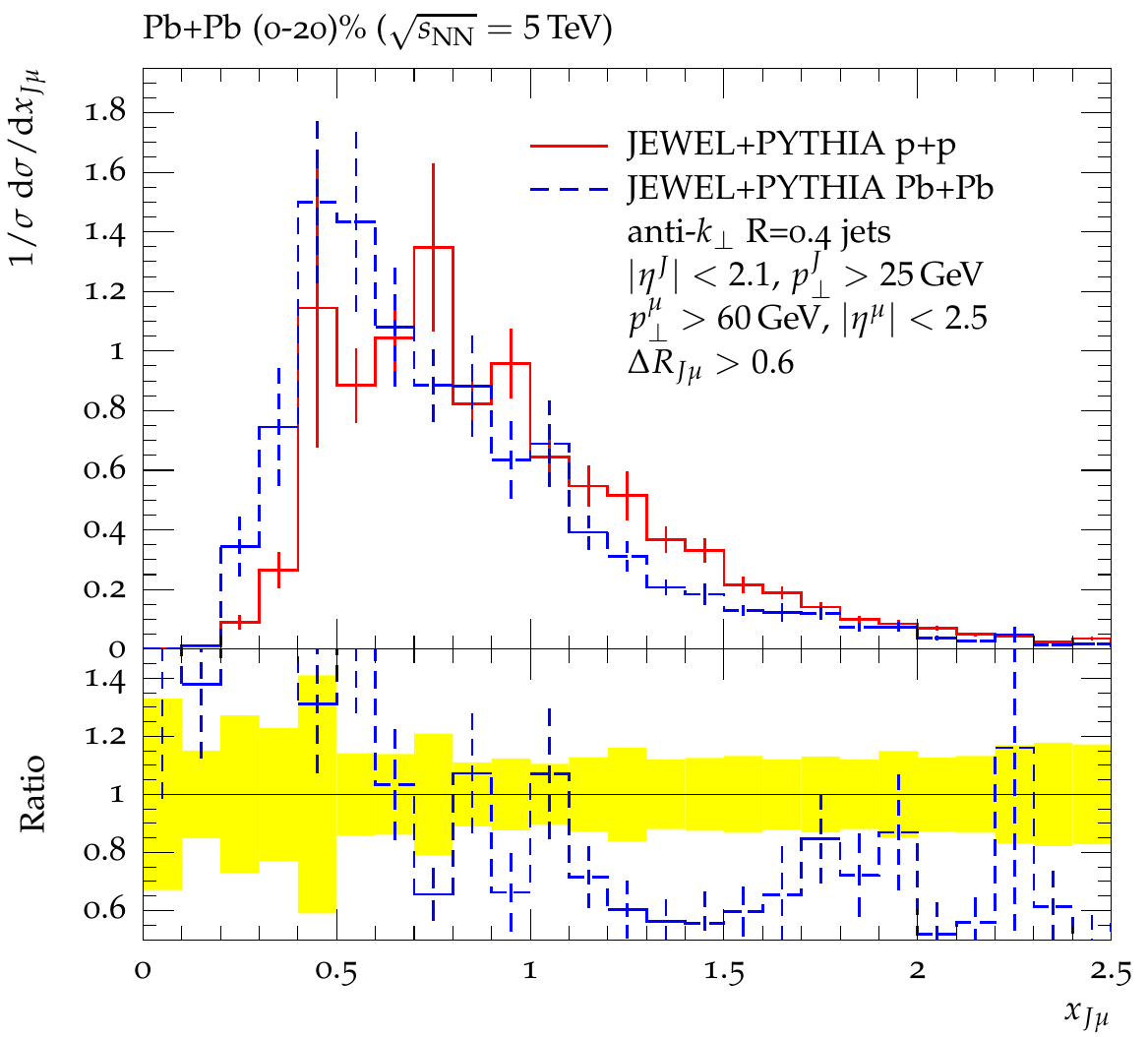} 
   \caption{Left: Azimuthal angle $\Delta \phi$ between the generator level $W$ and the jet in $W$+jet events compared to the azimuthal angle between the decay muon and the jet in central Pb+Pb events at $\sqrt{s_\text{NN}} = \unit[5.02]{TeV}$. Right: Momentum imbalance $x_{J\mu}$ with respect to the decay muon in $W$+jet events in p+p and central Pb+Pb collisions at $\sqrt{s_\text{NN}} = \unit[5.02]{TeV}$. In the ratio plots the dashed blue histogram is divided by the solid red one and the yellow band indicates the uncertainty on the latter.}
   \label{fig:wpjet_r4}
\end{figure*}

The left panel of Fig.~\ref{fig:wpjet_r4} shows the $\Delta \phi$ distributions in central ($0-20\%$) Pb+Pb events for the reconstructed jets with the generator level $W^{\pm}$ in the red line and with the leading lepton ($\mu$) in the event in the blue dotted line. We see that the $\Delta \phi$ distribution are similar for the $W^{\pm}$ and leading lepton and therefore we show the transverse momentum imbalance with the leptons. This is shown in the right panel of Fig.~\ref{fig:wpjet_r4} for p+p and central Pb+Pb collisions. Again, there is a clear shift towards larger asymmetries in central Pb+Pb events. 

\section{Conclusions}
We present an extension of \textsc{Jewel} with the additional capability of simulating $V$+jet events. This required only a slight modification of the framework to include the hard matrix elements for $V$+jet production. The jet quenching framework is independent of the underlying process (di-jet or $V$+jet), so that calculations for $V$+jet processes are performed without adjustments or tuning of parameters (the parameters of the background are obtained from a hydrodynamic calculation). This class of processes therefore constitutes an independent test of the predictivity of the \textsc{Jewel} framework.

Upon comparing with LHC Run\,I and Run\,II data we find generally good agreement for $\gamma/Z+$jet observables. This includes the shape, normalisation and boson $\pt$ dependence of the momentum imbalance, which shifts towards larger asymmetries in Pb+Pb events compared to p+p, and the $\Delta\phi$  distributions. The \textsc{Jewel} results fall below the data in the $\Delta \phi$ distributions for small angular separations in p+p and Pb+Pb. The same behaviour is observed in the angular distribution of di-jets~\cite{Zapp:2013vla}. This region is particularly sensitive to higher order corrections and it is thus likely that the discrepancy is caused by missing higher order matrix elements and a proper treatment of fragmentation photons in \textsc{Jewel}. The tendency to undershoot the region of very large $x_{JV}$ is probably also related to this. Nevertheless, the overall agreement between data and the \textsc{jewel} results is satisfactory and of similar quality as for other jet quenching observables, which showcases confidence in the jet quenching framework implemented in \textsc{Jewel} and its usability for performing predictions of jet observables and for comparisons with data. 

The theoretical understanding of jet quenching is not yet such that it can be used for reliable determination of medium parameters, but the successfull description of $V$+jet data by \textsc{Jewel} and other frameworks~\cite{Wang:1996pe,Dai:2012am,Ma:2013bia,Wang:2013cia,Casalderrey-Solana:2015vaa,Chien:2015hda} already tested against single-inclusive jet and di-jet data is a step towards a quantitative understanding of jet quenching.

\begin{acknowledgements}

We thank Guilherme Milhano for helpful comments on the manuscript and Chun Shen for providing the initial hydrodynamics parameters for our event generation at \unit[5]{TeV}. 
This work was supported by Funda\c{c}\~{a}o para a Ci\^{e}ncia e a Tecnologia (Portugal) under project CERN/FIS-NUC/0049/2015 and postdoctoral fellowship \\SFRH/BPD/102844/2014 (KCZ) and by the European Union as part of the FP7 Marie Curie Initial Training Network MCnetITN (PITN-GA-2012-315877). 
 
\end{acknowledgements}

%\bibliography{V+jet.bib}   % name your BibTeX data base

\begin{thebibliography}{}

%\cite{Aad:2015lcb}
\bibitem{Aad:2015lcb}
  G.~Aad {\it et al.} [ATLAS Collaboration],
  %``Centrality, rapidity and transverse momentum dependence of isolated prompt photon production in lead-lead collisions at $\sqrt{s_{\mathrm{NN}}} = 2.76$ TeV measured with the ATLAS detector,''
  Phys.\ Rev.\ C {\bf 93} (2016) no.3,  034914
  doi:10.1103/PhysRevC.93.034914
  [arXiv:1506.08552 [hep-ex]].
  %%CITATION = doi:10.1103/PhysRevC.93.034914;%%
  %11 citations counted in INSPIRE as of 31 May 2016

%\cite{Aad:2012ew}
\bibitem{Aad:2012ew}
  G.~Aad {\it et al.} [ATLAS Collaboration],
  %``Measurement of $Z$ boson Production in Pb+Pb Collisions at $\sqrt{s_{NN}}=2.76$ TeV with the ATLAS Detector,''
  Phys.\ Rev.\ Lett.\  {\bf 110} (2013) no.2,  022301
  doi:10.1103/PhysRevLett.110.022301
  [arXiv:1210.6486 [hep-ex]].
  %%CITATION = doi:10.1103/PhysRevLett.110.022301;%%
  %60 citations counted in INSPIRE as of 31 May 2016

%\cite{Aad:2014bha}
\bibitem{Aad:2014bha}
  G.~Aad {\it et al.} [ATLAS Collaboration],
  %``Measurement of the production and lepton charge asymmetry of $W$ bosons in Pb+Pb collisions at $\mathbf {\sqrt{\mathbf {s}_{\mathrm {\mathbf {NN}}}}=2.76\;TeV}$ with the ATLAS detector,''
  Eur.\ Phys.\ J.\ C {\bf 75} (2015) no.1,  23
  doi:10.1140/epjc/s10052-014-3231-6
  [arXiv:1408.4674 [hep-ex]].
  %%CITATION = doi:10.1140/epjc/s10052-014-3231-6;%%
  %21 citations counted in INSPIRE as of 31 May 20160907.4816

%\cite{Chatrchyan:2012vq}
\bibitem{Chatrchyan:2012vq}
  S.~Chatrchyan {\it et al.} [CMS Collaboration],
  %``Measurement of isolated photon production in $pp$ and PbPb collisions at $\sqrt{s_{NN}}=2.76$ TeV,''
  Phys.\ Lett.\ B {\bf 710} (2012) 256
  doi:10.1016/j.physletb.2012.02.077
  [arXiv:1201.3093 [nucl-ex]].
  %%CITATION = doi:10.1016/j.physletb.2012.02.077;%%
  %119 citations counted in INSPIRE as of 31 May 2016

%\cite{Chatrchyan:2014csa}
\bibitem{Chatrchyan:2014csa}
  S.~Chatrchyan {\it et al.} [CMS Collaboration],
  %``Study of Z production in PbPb and pp collisions at $ \sqrt{s_{\mathrm{NN}}}=2.76 $ TeV in the dimuon and dielectron decay channels,''
  JHEP {\bf 1503} (2015) 022
  doi:10.1007/JHEP03(2015)022
  [arXiv:1410.4825 [nucl-ex]].
  %%CITATION = doi:10.1007/JHEP03(2015)022;%%
  %20 citations counted in INSPIRE as of 31 May 2016

%\cite{Chatrchyan:2012nt}
\bibitem{Chatrchyan:2012nt}
  S.~Chatrchyan {\it et al.} [CMS Collaboration],
  %``Study of $W$ boson production in PbPb and $pp$ collisions at $\sqrt{s_{NN}}=2.76$ TeV,''
  Phys.\ Lett.\ B {\bf 715} (2012) 66
  doi:10.1016/j.physletb.2012.07.025
  [arXiv:1205.6334 [nucl-ex]].
  %%CITATION = doi:10.1016/j.physletb.2012.07.025;%%
  %91 citations counted in INSPIRE as of 31 May 2016


%\cite{Chatrchyan:2012gt}
\bibitem{Chatrchyan:2012gt}
  S.~Chatrchyan {\it et al.} [CMS Collaboration],
  %``Studies of jet quenching using isolated-photon+jet correlations in PbPb and $pp$ collisions at $\sqrt{s_{NN}}=2.76$ TeV,''
  Phys.\ Lett.\ B {\bf 718} (2013) 773
  doi:10.1016/j.physletb.2012.11.003
  [arXiv:1205.0206 [nucl-ex]].
  %%CITATION = doi:10.1016/j.physletb.2012.11.003;%%
  %150 citations counted in INSPIRE as of 31 May 2016

\bibitem{atlasypjet} CERN Report No. ATLAS-CONF-2012-121, 2012 (unpublished)

\bibitem{atlaszpjet} CERN Report No. ATLAS-CONF-2012-119, 2012 (unpublished)

% CMS y-jet update
\bibitem{cmsypjet} CERN Report No. CMS-PAS-HIN-13-006 (unpublished) 

\bibitem{cmszpjet} CERN Report No. CMS-PAS-HIN-15-013 (unpublished)	

% y-hadron stuff

%\cite{Nguyen:2010wb}
\bibitem{Nguyen:2010wb}
  M.~Nguyen,
  %``Jet Fragmentation in Vacuum and Medium with gamma-hadron Correlations in PHENIX,''
  J.\ Phys.\ Conf.\ Ser.\  {\bf 270} (2011) 012011
  doi:10.1088/1742-6596/270/1/012011
  [arXiv:1010.3187 [nucl-ex]].
  %%CITATION = doi:10.1088/1742-6596/270/1/012011;%%

%\cite{Hamed:2008yz}
\bibitem{Hamed:2008yz}
  A.~M.~Hamed [STAR Collaboration],
  %``gamma-hadron azimuthal correlations in STAR,''
  J.\ Phys.\ G {\bf 35} (2008) 104120
  doi:10.1088/0954-3899/35/10/104120
  [arXiv:0806.2190 [nucl-ex]].
  %%CITATION = doi:10.1088/0954-3899/35/10/104120;%%
  %14 citations counted in INSPIRE as of 01 Jun 2016


% y/Z-jet theory papers

%\cite{Wang:1996pe}
\bibitem{Wang:1996pe}
  X.~N.~Wang and Z.~Huang,
  %``Study medium induced parton energy loss in gamma + jet events of high-energy heavy ion collisions,''
  Phys.\ Rev.\ C {\bf 55} (1997) 3047
  doi:10.1103/PhysRevC.55.3047
  [hep-ph/9701227].
  %%CITATION = doi:10.1103/PhysRevC.55.3047;%%
  %143 citations counted in INSPIRE as of 01 Jun 2016

%\cite{Dai:2012am}
\bibitem{Dai:2012am}
  W.~Dai, I.~Vitev and B.~W.~Zhang,
  %``Momentum imbalance of isolated photon-tagged jet production at RHIC and LHC,''
  Phys.\ Rev.\ Lett.\  {\bf 110} (2013) no.14,  142001
  doi:10.1103/PhysRevLett.110.142001
  [arXiv:1207.5177 [hep-ph]].
  %%CITATION = doi:10.1103/PhysRevLett.110.142001;%%
  %47 citations counted in INSPIRE as of 08 Nov 2016

%\cite{Qin:2012gp}
\bibitem{Qin:2012gp}
  G.~Y.~Qin,
  %``Parton shower evolution in medium and nuclear modification of photon-tagged jets in Pb+Pb collisions at the LHC,''
  Eur.\ Phys.\ J.\ C {\bf 74} (2014) 2959
  doi:10.1140/epjc/s10052-014-2959-3
  [arXiv:1210.6610 [hep-ph]].
  %%CITATION = doi:10.1140/epjc/s10052-014-2959-3;%%
  %19 citations counted in INSPIRE as of 08 Nov 2016

%\cite{Ma:2013bia}
\bibitem{Ma:2013bia}
  G.~L.~Ma,
  %``Towards detailed tomography of high energy heavy-ion collisions by $\gamma$-jet,''
  Phys.\ Lett.\ B {\bf 724} (2013) 278
  doi:10.1016/j.physletb.2013.06.029
  [arXiv:1302.5873 [nucl-th]].
  %%CITATION = doi:10.1016/j.physletb.2013.06.029;%%
  %12 citations counted in INSPIRE as of 08 Nov 2016

%\cite{Wang:2013cia}
\bibitem{Wang:2013cia}
  X.~N.~Wang and Y.~Zhu,
  %``Medium Modification of $\gamma$-jets in High-energy Heavy-ion Collisions,''
  Phys.\ Rev.\ Lett.\  {\bf 111} (2013) no.6,  062301
  doi:10.1103/PhysRevLett.111.062301
  [arXiv:1302.5874 [hep-ph]].
  %%CITATION = doi:10.1103/PhysRevLett.111.062301;%%
  %42 citations counted in INSPIRE as of 08 Nov 2016

%\cite{Casalderrey-Solana:2015vaa}
\bibitem{Casalderrey-Solana:2015vaa}
  J.~Casalderrey-Solana, D.~C.~Gulhan, J.~G.~Milhano, D.~Pablos and K.~Rajagopal,
  %``Predictions for Boson-Jet Observables and Fragmentation Function Ratios from a Hybrid Strong/Weak Coupling Model for Jet Quenching,''
  JHEP {\bf 1603} (2016) 053
  doi:10.1007/JHEP03(2016)053
  [arXiv:1508.00815 [hep-ph]].
  %%CITATION = doi:10.1007/JHEP03(2016)053;%%
  %3 citations counted in INSPIRE as of 01 Jun 2016

%\cite{Chien:2015hda}
\bibitem{Chien:2015hda}
  Y.~T.~Chien and I.~Vitev,
  %``Towards the understanding of jet shapes and cross sections in heavy ion collisions using soft-collinear effective theory,''
  JHEP {\bf 1605} (2016) 023
  doi:10.1007/JHEP05(2016)023
  [arXiv:1509.07257 [hep-ph]].
  %%CITATION = doi:10.1007/JHEP05(2016)023;%%
  %8 citations counted in INSPIRE as of 01 Jun 2016

%\cite{Chang:2016gjp}
\bibitem{Chang:2016gjp}
  N.~B.~Chang and G.~Y.~Qin,
  %``Full jet evolution in quark-gluon plasma and nuclear modification of jet production and jet shape in Pb+Pb collisions at 2.76ATeV at the CERN Large Hadron Collider,''
  Phys.\ Rev.\ C {\bf 94} (2016) no.2,  024902
  doi:10.1103/PhysRevC.94.024902
  [arXiv:1603.01920 [hep-ph]].
  %%CITATION = doi:10.1103/PhysRevC.94.024902;%%
  %3 citations counted in INSPIRE as of 08 Nov 2016



%\cite{Wang:2010yz}
\bibitem{Wang:2010yz}
  X.~N.~Wang, H.~L.~Li, F.~M.~Liu, G.~L.~Ma and Y.~Zhu,
  %``Dihadron and gamma-hadron correlations from jet-induced medium excitation in high-energy heavy-ion collisions,''
  Nucl.\ Phys.\ A {\bf 855} (2011) 469
  doi:10.1016/j.nuclphysa.2011.02.108
  [arXiv:1012.2584 [nucl-th]].
  %%CITATION = doi:10.1016/j.nuclphysa.2011.02.108;%%
  %1 citations counted in INSPIRE as of 01 Jun 2016

%\cite{Renk:2009ur}
\bibitem{Renk:2009ur}
  T.~Renk,
  %``gamma-hadron correlations as a tool to trace the flow of energy lost from hard partons in heavy-ion collisions,''
  Phys.\ Rev.\ C {\bf 80} (2009) 014901
  doi:10.1103/PhysRevC.80.014901
  [arXiv:0904.3806 [hep-ph]].
  %%CITATION = doi:10.1103/PhysRevC.80.014901;%%
  %16 citations counted in INSPIRE as of 01 Jun 2016
 
%\cite{Zhang:2009fg}
\bibitem{Zhang:2009fg}
  H.~Zhang, J.~F.~Owens, E.~Wang and X.~N.~Wang,
  %``Gamma-Jet Tomography of Quark-Gluon Plasma in High-Energy Nuclear Collisions,''
  Nucl.\ Phys.\ A {\bf 830} (2009) 443C
  doi:10.1016/j.nuclphysa.2009.10.037
  [arXiv:0907.4816 [hep-ph]].
  %%CITATION = doi:10.1016/j.nuclphysa.2009.10.037;%%
  %6 citations counted in INSPIRE as of 01 Jun 2016
 
 
%\cite{Zapp:2012ak}
\bibitem{Zapp:2012ak}
K.~C.~Zapp, F.~Krauss and U.~A.~Wiedemann,
%``A perturbative framework for jet quenching,''
JHEP {\bf 1303} (2013) 080
[arXiv:1212.1599 [hep-ph]].
%%CITATION = ARXIV:1212.1599;%%
%17 citations counted in INSPIRE as of 09 May 2014


%\cite{CasalderreySolana:2012ef}
\bibitem{CasalderreySolana:2012ef}
  J.~Casalderrey-Solana, Y.~Mehtar-Tani, C.~A.~Salgado and K.~Tywoniuk,
  %``New picture of jet quenching dictated by color coherence,''
  Phys.\ Lett.\ B {\bf 725} (2013) 357
  doi:10.1016/j.physletb.2013.07.046
  [arXiv:1210.7765 [hep-ph]].
  %%CITATION = doi:10.1016/j.physletb.2013.07.046;%%
  %55 citations counted in INSPIRE as of 08 Nov 2016


%\cite{Floerchinger:2014yqa}
\bibitem{Floerchinger:2014yqa}
  S.~Floerchinger and K.~C.~Zapp,
  %``Hydrodynamics and Jets in Dialogue,''
  Eur.\ Phys.\ J.\ C {\bf 74} (2014) no.12,  3189
  doi:10.1140/epjc/s10052-014-3189-4
  [arXiv:1407.1782 [hep-ph]].
  %%CITATION = doi:10.1140/epjc/s10052-014-3189-4;%%
  %12 citations counted in INSPIRE as of 08 Nov 2016


\bibitem{bckgrnd}
 R.~Kunnawalkam Elayavalli and K.~C.~Zapp, in preparation 

%\cite{Zapp:2011ya}
\bibitem{Zapp:2011ya}
K.~C.~Zapp, J.~Stachel and U.~A.~Wiedemann,
%``A local Monte Carlo framework for coherent QCD parton energy loss,''
JHEP {\bf 1107} (2011) 118
[arXiv:1103.6252 [hep-ph]].
%%CITATION = ARXIV:1103.6252;%%
%21 citations counted in INSPIRE as of 10 Nov 2015

%\cite{Sjostrand:2006za}
\bibitem{Sjostrand:2006za}
T.~Sjostrand, S.~Mrenna and P.~Z.~Skands,
%``PYTHIA 6.4 Physics and Manual,''
JHEP {\bf 0605} (2006) 026
[hep-ph/0603175].
%%CITATION = HEP-PH/0603175;%%
%6850 citations counted in INSPIRE as of 10 Nov 2015


%\cite{Zapp:2013vla}
\bibitem{Zapp:2013vla}
K.~C.~Zapp,
%``JEWEL 2.0.0: directions for use,''
Eur.\ Phys.\ J.\ C {\bf 74} (2014) 2,  2762
[arXiv:1311.0048 [hep-ph]].
%9 citations counted in INSPIRE as of 09 Nov 2015
%%CITATION = ARXIV:1311.0048;%%

%\cite{Zapp:2013zya}
\bibitem{Zapp:2013zya}
K.~C.~Zapp,
%``Geometrical aspects of jet quenching in JEWEL,''
Phys.\ Lett.\ B {\bf 735} (2014) 157
[arXiv:1312.5536 [hep-ph]].
%%CITATION = ARXIV:1312.5536;%%
%9 citations counted in INSPIRE as of 09 Nov 2015


%\cite{Shen:2012vn}
\bibitem{Shen:2012vn}
C.~Shen and U.~Heinz,
%``Collision Energy Dependence of Viscous Hydrodynamic Flow in Relativistic Heavy-Ion Collisions,''
Phys.\ Rev.\ C {\bf 85} (2012) 054902
[Phys.\ Rev.\ C {\bf 86} (2012) 049903]
[arXiv:1202.6620 [nucl-th]].
%%CITATION = ARXIV:1202.6620;%%

\bibitem{Shen:2014vra} 
  C.~Shen, Z.~Qiu, H.~Song, J.~Bernhard, S.~Bass and U.~Heinz,
  %``The iEBE-VISHNU code package for relativistic heavy-ion collisions,''
  Comput.\ Phys.\ Commun.\  {\bf 199}, 61 (2016)
  doi:10.1016/j.cpc.2015.08.039
  [arXiv:1409.8164 [nucl-th]].
  %%CITATION = doi:10.1016/j.cpc.2015.08.039;%%


%\cite{Pumplin:2002vw}
\bibitem{Pumplin:2002vw}
  J.~Pumplin, D.~R.~Stump, J.~Huston, H.~L.~Lai, P.~M.~Nadolsky and W.~K.~Tung,
  %``New generation of parton distributions with uncertainties from global QCD analysis,''
  JHEP {\bf 0207} (2002) 012
  doi:10.1088/1126-6708/2002/07/012
  [hep-ph/0201195].
  %%CITATION = doi:10.1088/1126-6708/2002/07/012;%%
  %4628 citations counted in INSPIRE as of 08 Dec 2015


%\cite{Eskola:2009uj}
\bibitem{Eskola:2009uj}
  K.~J.~Eskola, H.~Paukkunen and C.~A.~Salgado,
  %``EPS09: A New Generation of NLO and LO Nuclear Parton Distribution Functions,''
  JHEP {\bf 0904} (2009) 065
  doi:10.1088/1126-6708/2009/04/065
  [arXiv:0902.4154 [hep-ph]].
  %%CITATION = doi:10.1088/1126-6708/2009/04/065;%%
  %540 citations counted in INSPIRE as of 08 Dec 2015

%\cite{Whalley:2005nh}
\bibitem{Whalley:2005nh}
  M.~R.~Whalley, D.~Bourilkov and R.~C.~Group,
  %``The Les Houches accord PDFs (LHAPDF) and LHAGLUE,''
  hep-ph/0508110.
  %%CITATION = HEP-PH/0508110;%%
  %446 citations counted in INSPIRE as of 08 Dec 2015

%\cite{Buckley:2010ar}
\bibitem{Buckley:2010ar}
A.~Buckley, J.~Butterworth, L.~Lonnblad, D.~Grellscheid, H.~Hoeth, J.~Monk, H.~Schulz and F.~Siegert,
%``Rivet user manual,''
Comput.\ Phys.\ Commun.\  {\bf 184} (2013) 2803
[arXiv:1003.0694 [hep-ph]].
%%CITATION = ARXIV:1003.0694;%%
%200 citations counted in INSPIRE as of 09 Nov 2015

%\cite{Cacciari:2008gp}
\bibitem{Cacciari:2008gp}
M.~Cacciari, G.~P.~Salam and G.~Soyez,
%``The Anti-k(t) jet clustering algorithm,''
JHEP {\bf 0804} (2008) 063
[arXiv:0802.1189 [hep-ph]].
%%CITATION = ARXIV:0802.1189;%%
%3131 citations counted in INSPIRE as of 09 Nov 2015

%\cite{Cacciari:2011ma}
\bibitem{Cacciari:2011ma}
M.~Cacciari, G.~P.~Salam and G.~Soyez,
%``FastJet User Manual,''
Eur.\ Phys.\ J.\ C {\bf 72} (2012) 1896
[arXiv:1111.6097 [hep-ph]].
%%CITATION = ARXIV:1111.6097;%%
%1053 citations counted in INSPIRE as of 09 Nov 2015
%35 citations counted in INSPIRE as of 09 Nov 2015


\end{thebibliography}


\end{document}